\documentclass{aastex631}

\definecolor{PineGreen}{RGB}{1, 121, 111}

\newcommand\G{\textcolor{cyan}}

\let\oldFootnote\footnote
\newcommand\nextToken\relax

\renewcommand\footnote[1]{%
\oldFootnote{#1}\futurelet\nextToken\isFootnote}

\newcommand\isFootnote{%
\ifx\footnote\nextToken\textsuperscript{,}\fi}

\usepackage{CJK}
\usepackage{hyperref}
\usepackage{epsfig}
\usepackage{color}
\usepackage{ulem}
\shorttitle{CSS CBs}
\shortauthors{Wang et al.}

\graphicspath{{./}{figures/}}

\begin{document}

\title{A Method of Rapidly Deriving Late-type Contact Binary Parameters and Its Application in the Catalina Sky Survey}

\correspondingauthor{Xu Ding}
\email{dingxu@ynao.ac.cn}

\author[0009-0002-1546-8442]{JinLiang Wang}
\affiliation{Yunnan Observatories, Chinese Academy of Sciences (CAS), P.O. Box 110, 650216 Kunming, People’s Republic of China}
\affiliation{Key Laboratory of the Structure and Evolution of Celestial Objects, Chinese Academy of Sciences, P.O. Box 110, 650216 Kunming, People’s Republic of China}
\affiliation{Center for Astronomical Mega-Science, Chinese Academy of Sciences, 20A Datun Road, Chaoyang District, Beijing 100012, People’s Republic of China}
\affiliation{University of the Chinese Academy of Sciences, Yuquan Road 19\#, Shijingshan Block, 100049 Beijing, People’s Republic of China}

\author{Xu Ding}
\affiliation{Yunnan Observatories, Chinese Academy of Sciences (CAS), P.O. Box 110, 650216 Kunming, People’s Republic of China}
\affiliation{Key Laboratory of the Structure and Evolution of Celestial Objects, Chinese Academy of Sciences, P.O. Box 110, 650216 Kunming, People’s Republic of China}
\affiliation{Center for Astronomical Mega-Science, Chinese Academy of Sciences, 20A Datun Road, Chaoyang District, Beijing 100012, People’s Republic of China}
\affiliation{University of the Chinese Academy of Sciences, Yuquan Road 19\#, Shijingshan Block, 100049 Beijing, People’s Republic of China}

\author{JiaJia Li}
\affiliation{Yunnan Observatories, Chinese Academy of Sciences (CAS), P.O. Box 110, 650216 Kunming, People’s Republic of China}
\affiliation{Key Laboratory of the Structure and Evolution of Celestial Objects, Chinese Academy of Sciences, P.O. Box 110, 650216 Kunming, People’s Republic of China}
\affiliation{Center for Astronomical Mega-Science, Chinese Academy of Sciences, 20A Datun Road, Chaoyang District, Beijing 100012, People’s Republic of China}
\affiliation{University of the Chinese Academy of Sciences, Yuquan Road 19\#, Shijingshan Block, 100049 Beijing, People’s Republic of China}

\author{JianPing Xiong}
\affiliation{Yunnan Observatories, Chinese Academy of Sciences (CAS), P.O. Box 110, 650216 Kunming, People’s Republic of China}

\author{QiYuan Cheng}
\affiliation{Yunnan Observatories, Chinese Academy of Sciences (CAS), P.O. Box 110, 650216 Kunming, People’s Republic of China}
\affiliation{Key Laboratory of the Structure and Evolution of Celestial Objects, Chinese Academy of Sciences, P.O. Box 110, 650216 Kunming, People’s Republic of China}
\affiliation{Center for Astronomical Mega-Science, Chinese Academy of Sciences, 20A Datun Road, Chaoyang District, Beijing 100012, People’s Republic of China}
\affiliation{University of the Chinese Academy of Sciences, Yuquan Road 19\#, Shijingshan Block, 100049 Beijing, People’s Republic of China}

\author{KaiFan Ji}
\affiliation{Yunnan Observatories, Chinese Academy of Sciences (CAS), P.O. Box 110, 650216 Kunming, People’s Republic of China}
\affiliation{Key Laboratory of the Structure and Evolution of Celestial Objects, Chinese Academy of Sciences, P.O. Box 110, 650216 Kunming, People’s Republic of China}
\affiliation{Center for Astronomical Mega-Science, Chinese Academy of Sciences, 20A Datun Road, Chaoyang District, Beijing 100012, People’s Republic of China}
\affiliation{University of the Chinese Academy of Sciences, Yuquan Road 19\#, Shijingshan Block, 100049 Beijing, People’s Republic of China}

\begin{abstract}

With the continuous development of large optical surveys, a large number of light curves of late-type contact binary systems (CBs) have been released. Deriving parameters for CBs using the the WD program and the PHOEBE program poses a challenge. Therefore, this study developed a method for rapidly deriving light curves based on the Neural Networks (NN) model combined with the Hamiltonian Monte Carlo (HMC) algorithm (NNHMC). The neural network was employed to establish the mapping relationship between the parameters and the pregenerated light curves by the PHOEBE program, and the HMC algorithm was used to obtain the posterior distribution of the parameters. The NNHMC method was applied to a large contact binary sample from the Catalina Sky Survey, and a total of 19,104 late-type contact binary parameters were derived. Among them, 5172 have an inclination greater than $70^\circ$ and a temperature difference less than 400 K. The obtained results were compared with the previous studies for 30 CBs, and there was an essentially consistent goodness-of-fit ($R^2$) distribution between them. The NNHMC method possesses the capability to simultaneously derive parameters for a vast number of targets. Furthermore, it can provide an extremely efficient tool for rapid derivation of parameters in future sky surveys involving large samples of CBs.

\end{abstract}


\section{Introduction} \label{sec:intro}

Late-type contact binary systems (CBs), also referred to as W Ursae Majoris (W UMa) variables, are eclipsing binaries where both components have filled their Roche lobes. They share a common envelope \citep{1979ApJ...231..502L}, and most of them have similar temperatures \citep{1941ApJ....93..133K}. Due to the contact nature of CBs, mass and energy can be exchanged between the components \citep{1959cbs..book.....K,1968ApJ...151.1123L,1968ApJ...153..877L}, leading to variations in their light curves. Thus, investigating the variations in the light curve is a powerful method for studying the formation and evolution of CBs. 

The existence of an evolutionary sequence among different types of CBs is still under ongoing research. Currently, only a limited number of CBs have been studied \citep{2001MNRAS.328..635Q,2005ApJ...629.1055Y,2013MNRAS.430.2029Y,2021ApJS..254...10L,2022MNRAS.510.5315P,2023PASP..135g4202W,2024MNRAS.527.6406L}. The large sample size is crucial for constraining the evolution model of CBs \citep{2006AcA....56..199S} and its angular momentum loss properties and nuclear evolution paths, especially the possible effects of these characteristics and paths on the orbital periods \citep{2016ApJ...832..138C,2020MNRAS.492.2731J} and evolution products of different types of CBs \citep{2015AJ....150...69Y,2019MNRAS.485.4588L}.

With the development of various sky survey telescopes, such as the Kepler mission \citep{borucki2010kepler,2010ApJ...713L..79K}, the Asteroid Terrestrial-impact Last Alert System \citep[ATLAS;][]{2018AJ....156..241H}, and the Catalina Sky Survey \citep[CSS;][]{2017MNRAS.465.4678M}, a large number of light curves were being released sequentially. Researchers have constructed a large sample of CBs for further statistical studies \citep{1995ApJ...446L..19R,2011A&A...528A..90N,2017MNRAS.465.4678M}. However, previous studies involving large samples of CBs were primarily limited to the investigation of their light curve morphology, such as period and amplitude, without deriving stellar parameters. Currently, the commonly used methods for deriving parameters of CBs involve the Wilson-Devinney (WD) program \citep{1990ApJ...356..613W,2012AJ....144...73W,1971ApJ...166..605W,2010ApJ...723.1469W,2007ApJ...661.1129V} and the PHysics Of Eclipsing BinariEs (PHOEBE) program \citep{2016ApJS..227...29P,2019JOSS....4.1864F}. However, deriving parameters for a target using the WD program and the PHOEBE program typically requires several hours to several days. Therefore, for large samples of CBs, rapidly deriving parameters for all targets remains a significant challenge. The use of machine learning methods can accelerate the derivation of parameters for detached binaries \citep{2008ApJ...687..542P} and contact binaries \citep{2021PASJ...73..786D}. However, these methods do not inherently provide corresponding parameter uncertainties. Although using machine learning with the Markov Chain Monte Carlo (MCMC) algorithm \citep{2022AJ....164..200D} can provide corresponding parameter uncertainties, it can also quickly derive the parameters of a single light curve. However, due to its algorithmic limitations, it is not suitable for deriving parameters for a large number of targets simultaneously.

Due to the fact that generating a synthetic light curve with the PHOEBE program takes several seconds to tens of seconds, subsequently using the MCMC algorithm to derive parameters for a target can take several hours to several days \citep{2019JOSS....4.1864F}. In this study, the JAX framework \citep{jax2018github} was used to construct a neural network (NN) to replace the PHOEBE program in generating synthetic light curves, and the use of the NN model combined with the Hamiltonian Monte Carlo (HMC) algorithm \citep{MD2011The} allows for the rapid derivation of parameters for a target. Furthermore, it provides uncertainties for all parameters. This paper applies the NNHMC method to a large CBs sample from CSS Data Release 2 
\citep[CSDR2 \footnote{\url{http://nesssi.cacr.caltech.edu/DataRelease/}};][]{2014ApJS..213....9D}, a total of 19,104 CB parameters were obtained, of which 5172 have an inclination greater than $70^\circ$ and a temperature difference less than 400 K.

This study focuses on the rapid derivation of parameters for CBs observed in the CSS and provides a corresponding star catalog. In Section~\ref{sec:CBs candidates}, a brief introduction is provided for the CSS data used in this study, along with some data preprocessing steps. In Section~\ref{sec:nnhmc}, a detailed description is given of the NN model combined with the HMC algorithm, including model establishment, evaluation, and analysis. In Section~\ref{sec:result and discussion}, the parameters of CBs in the CSS are obtained and discussed. And conclusions are presented in Section~\ref{sec:Conclusion}.

\section{Data} \label{sec:CBs candidates}

This article's experimental samples were selected from CSDR2 \citep{2014ApJS..213....9D}. The CSS began in 2004, covering the sky between declinations of -75°and +65°using three telescopes. To maximize throughput, all observation results were unfiltered, and the data was then transformed to the V-band \citep{2013ApJ...763...32D}. The orbital periods of most CBs typically range from 0.25 days to 0.5 days, and the typical photometric uncertainty falls within the range of $0.05$ mag to $0.10$ mag.

\cite{2014ApJS..213....9D} identified 30,743 CBs (EW-type stars) in the CSS. In this study, the initial selection excluded 2335 CBs with parameters derived by \cite{2020ApJS..247...50S} and 30 CBs with parameters derived by \cite{2022MNRAS.512.1244C}. Then, to obtain an initial estimate of the effective temperature of the primary star in the CBs, this study cross-referenced the sample with Gaia DR2 data\citep{2018A&A...616A..10G}, resulting in 21,469 CBs with Gaia-measured temperatures, leaving a remaining 19,104 CBs. Finally, to standardize the initial sample to the same scale as the synthetic light curves generated by the PHOEBE program \citep{2016ApJS..227...29P,2019JOSS....4.1864F}, preprocessing was performed on the initial sample. The first step involved using the Isolation Forest anomaly detection algorithm \citep{8888179} to remove outliers from the original light curves. Subsequently, the Lomb-Scargle method \citep{1976Ap&SS..39..447L,1982ApJ...263..835S} combined with bootstrap method \citep{efron1979individual} were employed to calculate the period of the original light curves and fold them. Finally, each light curve underwent zero-mean and zero-phase processing. In the end, the curve from the initial MJD-mag curve to the single-period phase-mag curve after preprocessing was obtained.

\section{NN Model combined with HMC Algorithm (NNHMC)} \label{sec:nnhmc}

Because it takes several hours to several days to derive the parameters of a light curve by using the PHOEBE program combined with the MCMC algorithm \citep{2019JOSS....4.1864F}, it is difficult to meet the needs of a large-sample sky survey. Therefore, this study uses a neural network to replace the PHOEBE program in generating synthetic light curves to expedite the derivation process of the MCMC algorithm. The process of the proposed method involves first establishing a NN model from parameters to a synthetic light curve. Then, using the observed light curve and primary star temperature as input, the NN model and the HMC algorithm were combined to quickly derive the light curve. Finally, the parameters and their uncertainties for the observed light curve were obtained.
\subsection{Establishment and Analysis of the Neural Network}\label{sec:training process}

\begin{figure*}
	\centering
	\includegraphics[scale=0.6]{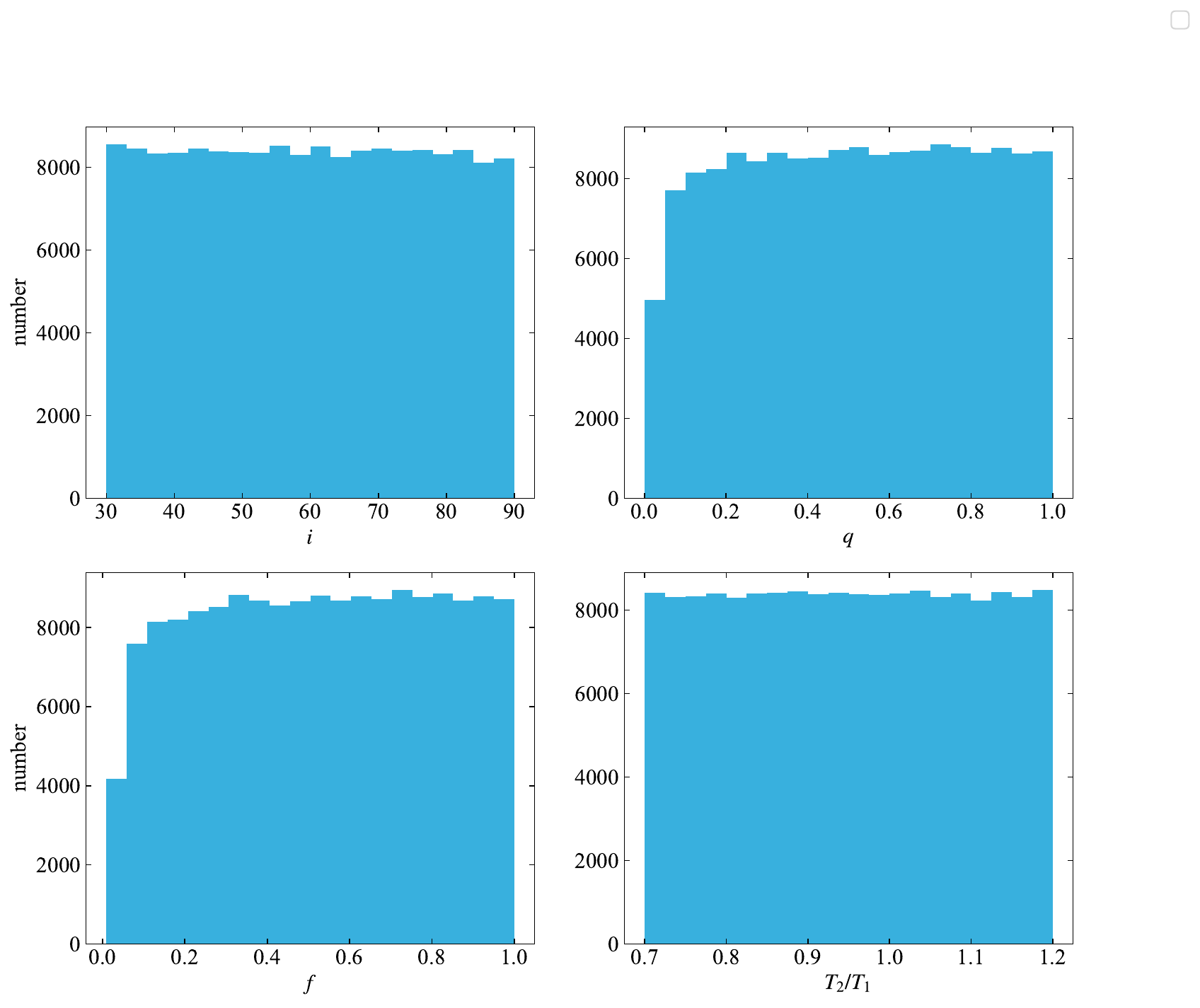}
    \caption{The distribution of the parameters of the sample set.}
    \label{fig: parameter_distribution}
\end{figure*}

\begin{figure*}
	\centering
	\includegraphics[scale=0.4]{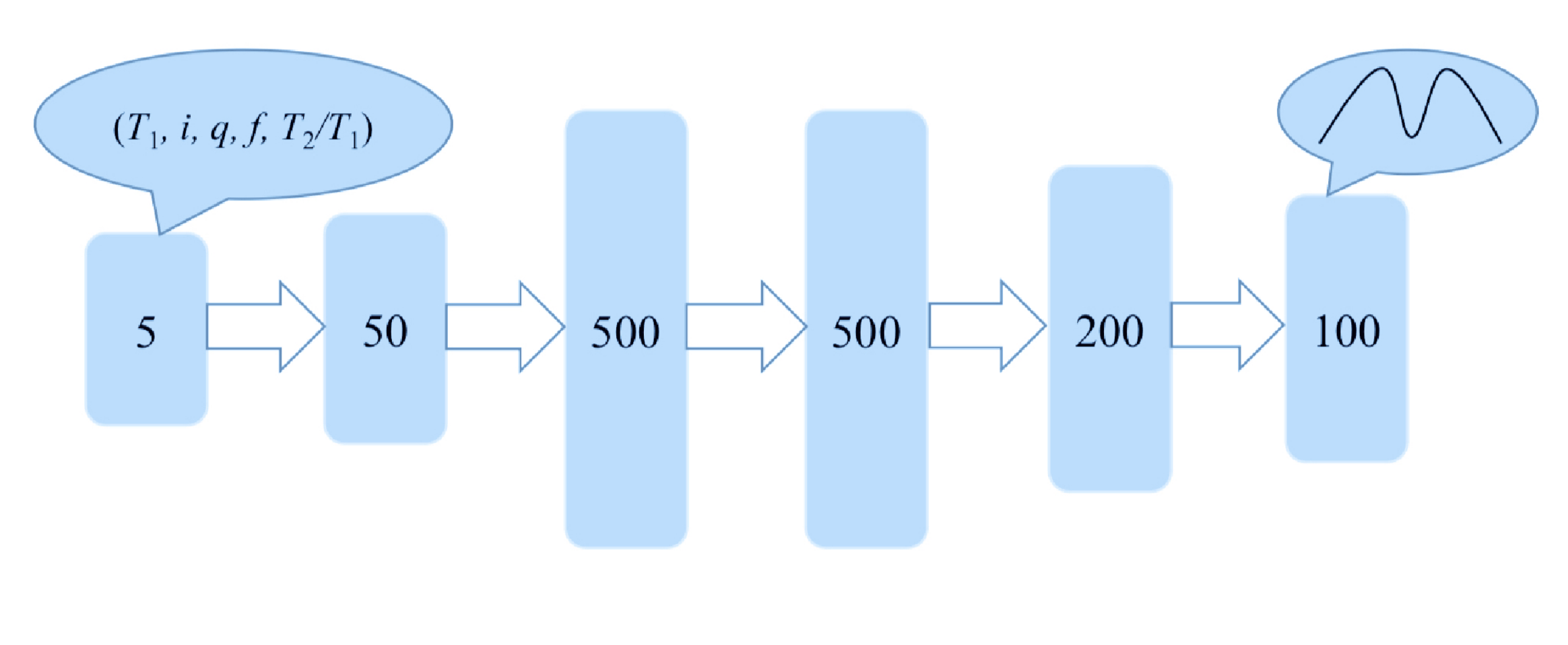}
    \caption{The structure of the neural network: 5 input parameters, 4 hidden layers, the output is a light curve with 100 phase points, the number in the box represents the number of nodes in each layer.}
    \label{fig: block_diagram}
\end{figure*}

Establishing a NN model from parameters to synthetic light curves requires a large number of samples. Therefore, this study uses the PHOEBE program to generate 170,000 synthetic light curves of CBs for training, validation, and testing of the neural network. The temperature of the primary star ($T_1$), the effective temperature of the secondary star ($T_2$), the orbital inclination ($i$), the mass ratio ($q$), the fill-out factor ($f$), and the passband are the main parameters influencing the light curves of CBs. This study generates synthetic light curves for CBs using the PHOEBE program based on these parameters. The passband was set to Johnson:V, $T_1$ was set between 4000K and 8000K, $i$ was set between 30° and 90°, $q$ was set between 0 and 1, $f$ was set between 0 and 1, and $T_2/T_1$ was set between 0.7 and 1.2, essentially covering the CSS. The additional parameter settings in the PHOEBE program were referenced in \cite{2022AJ....164..200D}. The histogram of parameter statistics corresponding to the 170,000 synthetic light curves generated by the PHOEBE program is shown in Figure~\ref{fig: parameter_distribution}. The sample set is basically flat on all sampled parameters making it an effectively even grid. There are less models of small $q$ and $f$ because the systems are significantly less likely to be in contact due to physical limitations. JAX is a framework developed by Google Research for high-performance numerical computing and machine learning research \citep{jax2018github}. It provides a Just-In-Time (JIT) component for capturing code and optimizing it for the Accelerated Linear Algebra (XLA) compiler, significantly improving the performance of TensorFlow and PyTorch. Flax is an open-source neural network library on JAX \citep{flax2020github}. In this paper, JAX combined with Flax was used to construct the neural network (multi-layer perceptron) of parameter to light curve, which was used to generate synthetic light curves instead of the PHOEBE program, and improve the deriving speed of the HMC algorithm.  The framework of the neural network is illustrated in Figure~\ref{fig: block_diagram}.

As shown in Figure~\ref{fig: block_diagram}, the input of the neural network model is 5 parameters, namely $T_1$, $i$, $q$, $f$, and $T_2/T_1$, and the output is a light curve with 100 phase points, with 4 hidden layers in the middle, and the activation function of each layer is Rectified Linear Unit (Relu) \citep{nair2010rectified}. The purpose of introducing the activation function is to increase the nonlinear fitting ability of the neural network. The synthetic light curves of 170,000 CBs generated by the PHOEBE program were taken as samples, of which 150,000 were training sets, 10,000 were verification sets and 10,000 were test sets. The loss function used in the model training and validation process was the mean-square error (MSE) between the predicted value and the true value. On a computer with 16GB of random-access memory and a 3.40 GHz 16-core CPU, the trained model can generate an average of 100-point light curves in just $0.001$s. In comparison, under the same conditions, the PHOEBE program takes an average of 5s to generate 100-point light curves. The speed of generating light curves by the model was significantly faster than that of the PHOEBE program. To evaluate the trained neural network model, this model was used to predict 10,000 light curves in the test set, and the standard deviation of the difference between the predicted light curve of each set of parameters and the real light curve corresponding to the parameters was calculated, which was often used to characterize the accuracy of the neural network model.

\begin{figure}
	\centering
	\includegraphics[scale=0.6]{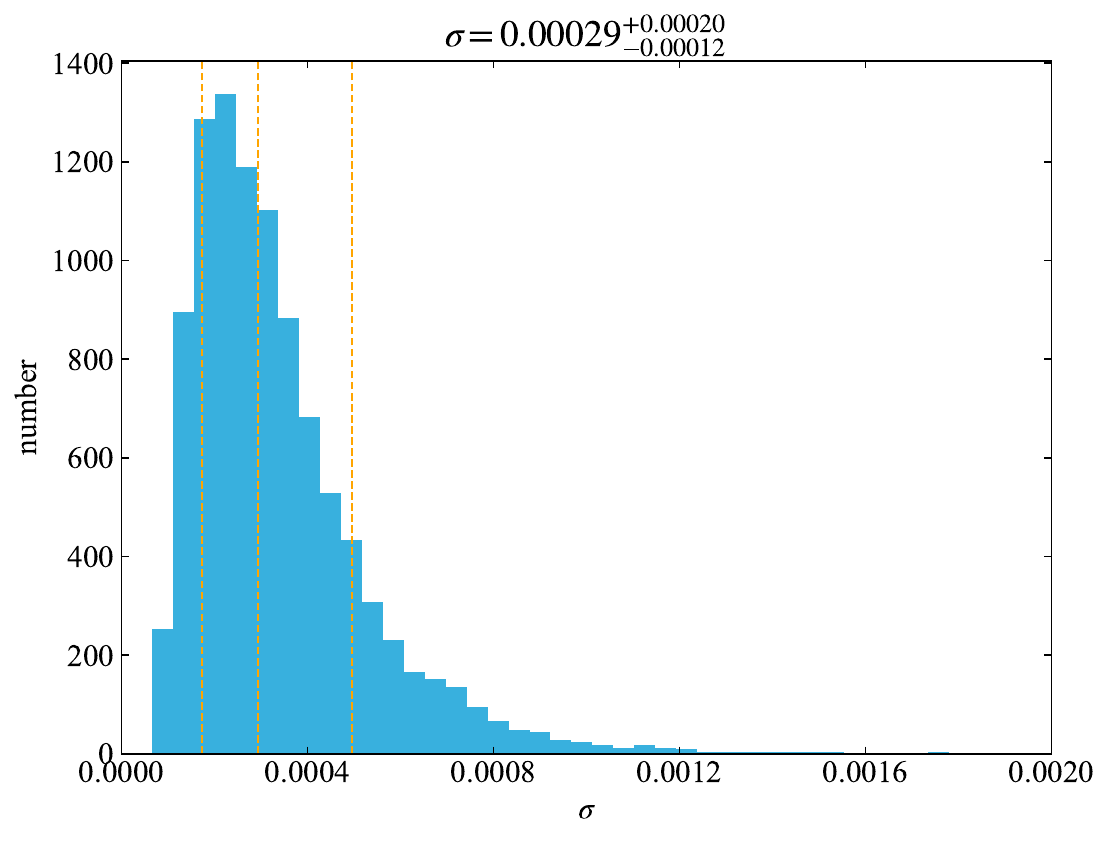}
    \caption{The distribution of the standard deviation of residuals between 10,000 synthetic light curves and the light curves predicted by a neural network in the test set. The three orange vertical dashed lines in the figure represent the lower error at the 16 percentile, the median at the 50 percentile, and the upper error at the 84 percentile of the distribution, respectively.}
    \label{fig: theory_rmse}
\end{figure}

\begin{figure*}
	\centering
	\includegraphics[scale=0.5]{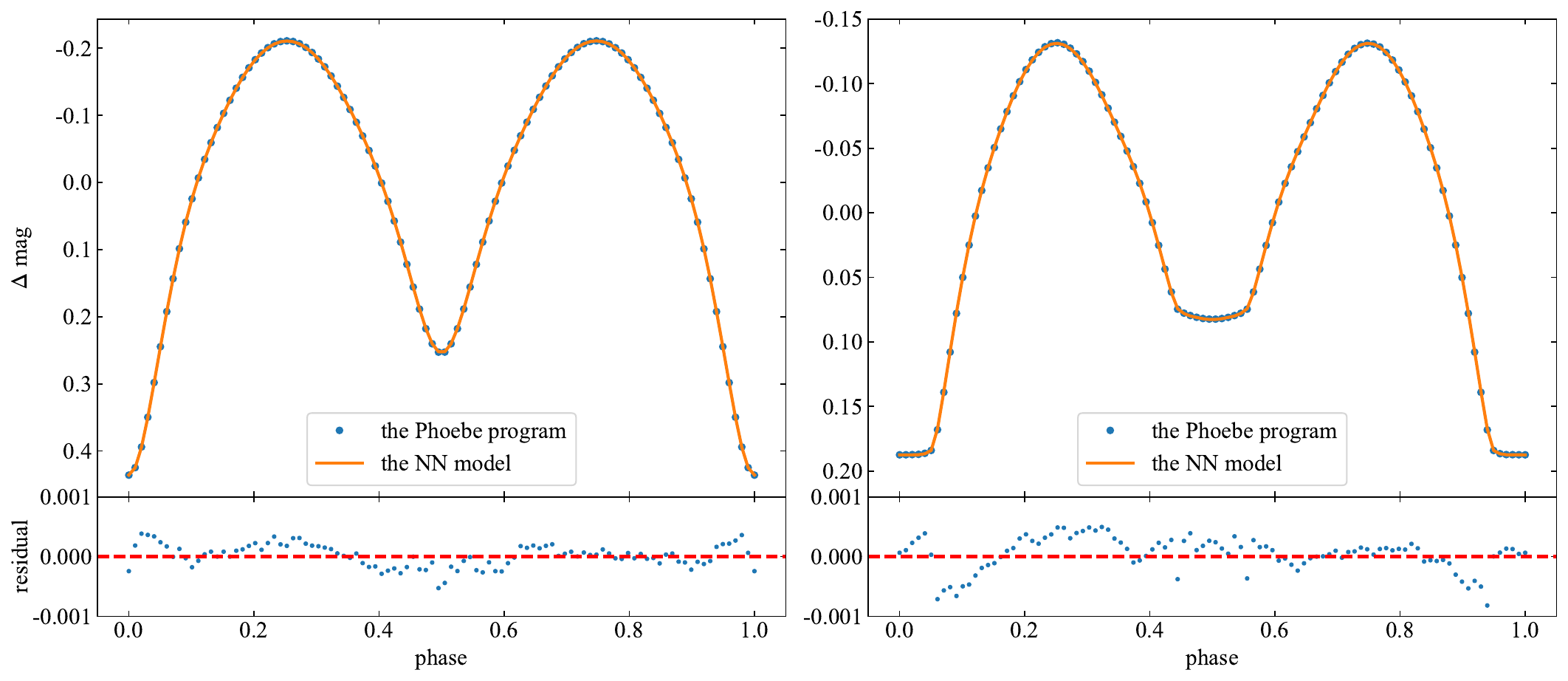}
    \caption{The proximity of the two light curves generated by the two methods in the test set. The upper part of the figure displays the blue light curve generated by the PHOEBE program, while the orange light curve represents that generated by the NN model. In the lower half of the figure, blue dots represent residuals between these two light curves (orange minus blue), with a dashed red line indicating zero value.}
    \label{fig: theory_predicted}
\end{figure*}

Figure~\ref{fig: theory_rmse} displays the distribution of the standard deviation of residuals between 10,000 synthetic light curves and the light curves predicted by a neural network in the test set. The three orange vertical dashed lines in the figure represent the lower error at the 16 percentile, the median at the 50 percentile, and the upper error at the 84 percentile of the distribution, respectively. It can be observed that the main distribution lies within $0.00029^{+0.00020}_{-0.00012}$, with the maximum not exceeding $0.002$, meeting the precision requirements. The comparison between the light curve generated by the NN model and the light curve generated by the PHOEBE program in the test set is illustrated in Figure~\ref{fig: theory_predicted}. The upper part of the figure displays the blue light curve generated by the PHOEBE program, while the orange light curve represents that generated by the NN model. In the lower half of the figure, blue dots represent residuals between these two light curves (orange minus blue), with a dashed red line indicating zero value. For the upper left light curve, the corresponding parameters were $T_1 = 5275$ K, $i = 76.17$°, $q = 0.70$, $f = 0.37$, and $T_2/T_1 = 0.917$, with a residual standard deviation of $0.00018$. Similarly, for the upper right light curve, the corresponding parameters were $T_1 = 4613$ K, $i = 87.81$°, $q = 0.15$, $f = 0.56$, and $T_2/T_1 = 0.883$, with a residual standard deviation of $0.00029$. It can be observed that both methods generate highly similar light curves.

\subsection{HMC algorithm}\label{sec:hmc}

Traditional MCMC algorithms typically employ the Metropolis-Hastings (M-H) algorithm \citep{2019JOSS....4.1864F,2022AJ....164..200D}. One issue with this method is that it often exhibits random walk behavior during the sampling process. However, random walks are not highly efficient for exploring parameter space, as the average exploration distance is proportional to the square root of the number of iterations. Simply increasing the step size can reduce acceptance rates. Another issue is sensitivity to the initial values. The HMC algorithm provides a novel approach by introducing an additional dimension, allowing the sampling trajectory to advance along iso-density lines (Hamiltonian constant) \citep{MD2011The,bingham2019pyro,phan2019composable}. The significance of introducing an additional dimension lies in the ability to achieve leaps, meaning one can easily move from point A to point B without being constrained by the intricacies of the original space. This approach not only enhances the exploration efficiency of the parameter space but also maintains a relatively high acceptance rate. Simultaneously, it exhibits insensitivity to the initial values.

\subsection{Performance of the NNHMC method}\label{sec:performance}

\begin{figure}
	\centering
	\includegraphics[scale=0.6]{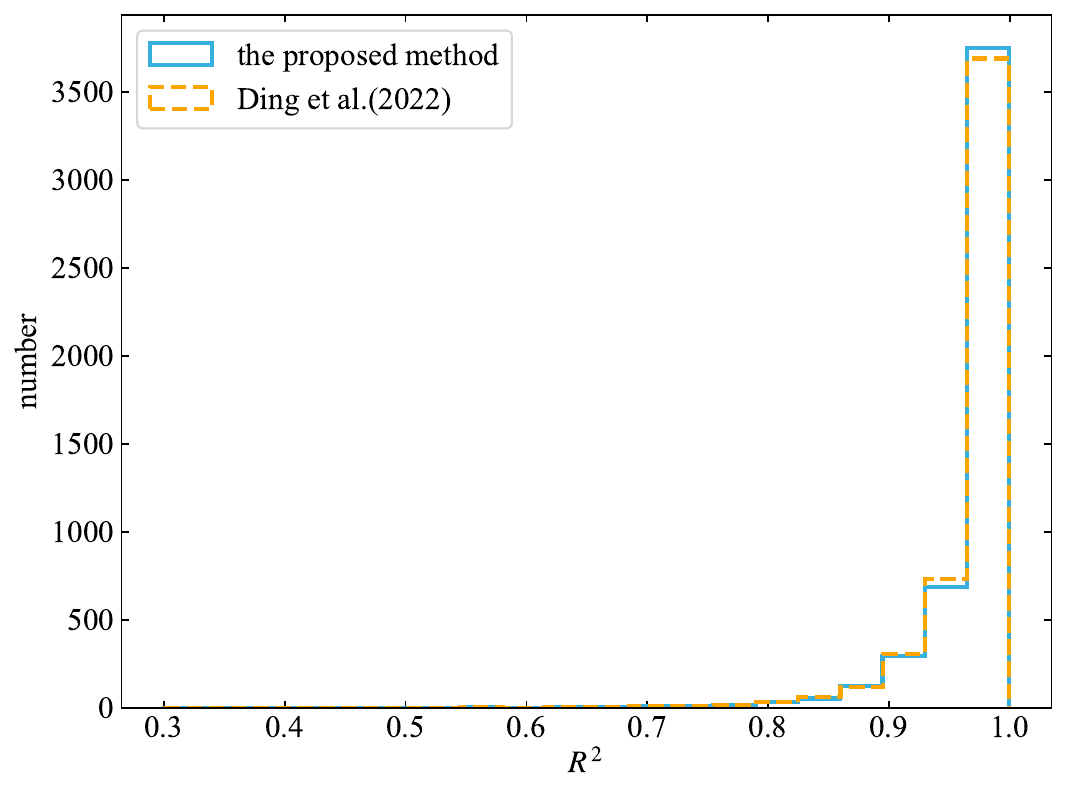}
    \caption{For 5000 randomly selected synthetic light curves (adding $0.01$ mag Gaussian noise) from the test set, their parameters were derived by \cite{2022AJ....164..200D} and the proposed method respectively, and the parameters obtained by the two methods were input into the PHOEBE program to calculate the goodness-of-fit ($R^2$) between the newly generated light curve and the actual light curve respectively, where the blue solid bar shows the $R^2$ distribution curve obtained by the proposed method, and the orange dashed bar indicates the $R^2$ distribution curve obtained by \cite{2022AJ....164..200D}.}
    \label{fig: hmc_Ding_r2}
\end{figure}

To verify the high efficiency of the proposed method, the NN model combined with the MCMC algorithm \citep{2022AJ....164..200D} and the NNHMC method were used to derive the parameters of the above 5000 synthetic light curves with $0.01$ magnitude noise added, and the derived parameters obtained were input into the PHOEBE program, respectively. The goodness-of-fit ($R^2$) between the newly generated light curves and the actual light curves was calculated, and it was used as an evaluation index of the derived parameters. The range of $R^2$ is between (0, 1), and the closer $R^2$ is to 1, the better the derivation of the parameters. The distribution of $R^2$ corresponding to 5000 light curves is shown in Figure~\ref{fig: hmc_Ding_r2}. The blue solid bar shows the $R^2$ distribution curve obtained by the proposed method, and the orange dashed bar indicates the $R^2$ distribution curve obtained by \cite{2022AJ....164..200D}. It can be seen that the $R^2$ values of the two methods were basically greater than 0.8, and the distribution was basically the same. Hypothesis testing \citep{raftery1995hypothesis} is done on both distributions, and the test results are consistent, indicating that the two methods are essentially the same in terms of accuracy. However, on a computer with 16G random access memory and a 3.40GHz main-frequency 16-core CPU, the time to derive a light curve using the NN model combined with the MCMC algorithm \citep{2022AJ....164..200D} was about 17s, while the time to derive a light curve using the NNHMC method was about 0.13s, and the computation time was shortened by 2 orders of magnitude. It was very suitable for parameter derivation of large sample CBs.

\section{Result and discussion} \label{sec:result and discussion}

In this paper, the proposed method was used to quickly derive 19,104 CBs from the CSS \citep{2014ApJS..213....9D}, and its parameters were obtained. At the same time, in order to obtain reliable parameters, we applied additional selection criteria to the obtained CB parameters. These selection criteria include three aspects: First, the obtained parameters were inputted into the PHOEBE program to generate a new light curve. Subsequently, a direct comparison was made between the original light curve and the generated one in order to exclude any samples that did not exhibit congruence. Second, to mitigate the occurrence of degeneracy, this study excludes CBs with an inclination below $70^\circ$. \cite{2013CoSka..43...27H} pointed out that photometric light curves are not an effective tool for analyzing low-inclination systems because the number of similar (i.e., degenerate) light curves increases as the inclination decreases. From a light curve standpoint, the distinction between ellipsoidal variables and CBs is fairly small at low inclinations, so it might be difficult to distinguish the difference between contact and ellipsoidal variation. For CBs whose orbital plane has a large tilt relative to the observer ($i > 70^\circ$), it is characterized by the possibility of large variations in brightness due to orbital eclipses. Third, due to a limitation of the PHOEBE program itself, this study removed CBs with temperature differences greater than 400 K. \cite{2018arXiv180408781K} pointed out that there are inherent flaws in using the  PHOEBE program to model CBs. These are negligible for CBs with similar temperatures. However, as the temperature difference becomes more pronounced, there is a clear jump in the neck region of the temperature distribution of CBs. There are significant differences in the light curves produced with smoothed temperatures in the localized neck region. Therefore, this paper finally obtains the parameter list of 5172 CBs in the CSS, and all the derived parameters are shown in Table~\ref{Tab:1}.

\begin{figure}
	\centering
	\includegraphics[scale=0.6]{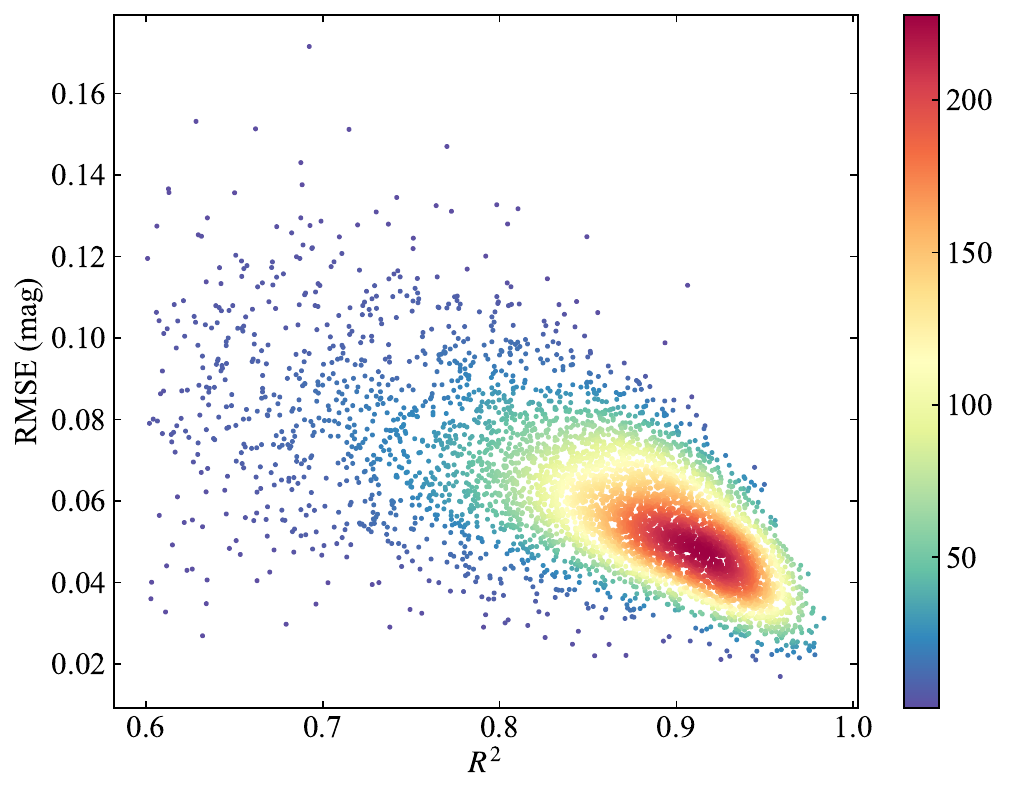}
    \caption{The parameters of 5172 CBs in the CSS were derived using the NNHMC method, and these obtained parameters were subsequently utilized as input for the PHOEBE program to calculate the $R^2$ and RMSE values between the newly generated light curve and the observed light curve.
    The distribution of $R^2$ and RMSE is shown. Colors represent the number of objects in each bin.}
    \label{fig: r2_rmse_distribution}
\end{figure}

\begin{figure*}
	\centering
	\includegraphics[scale=0.5]{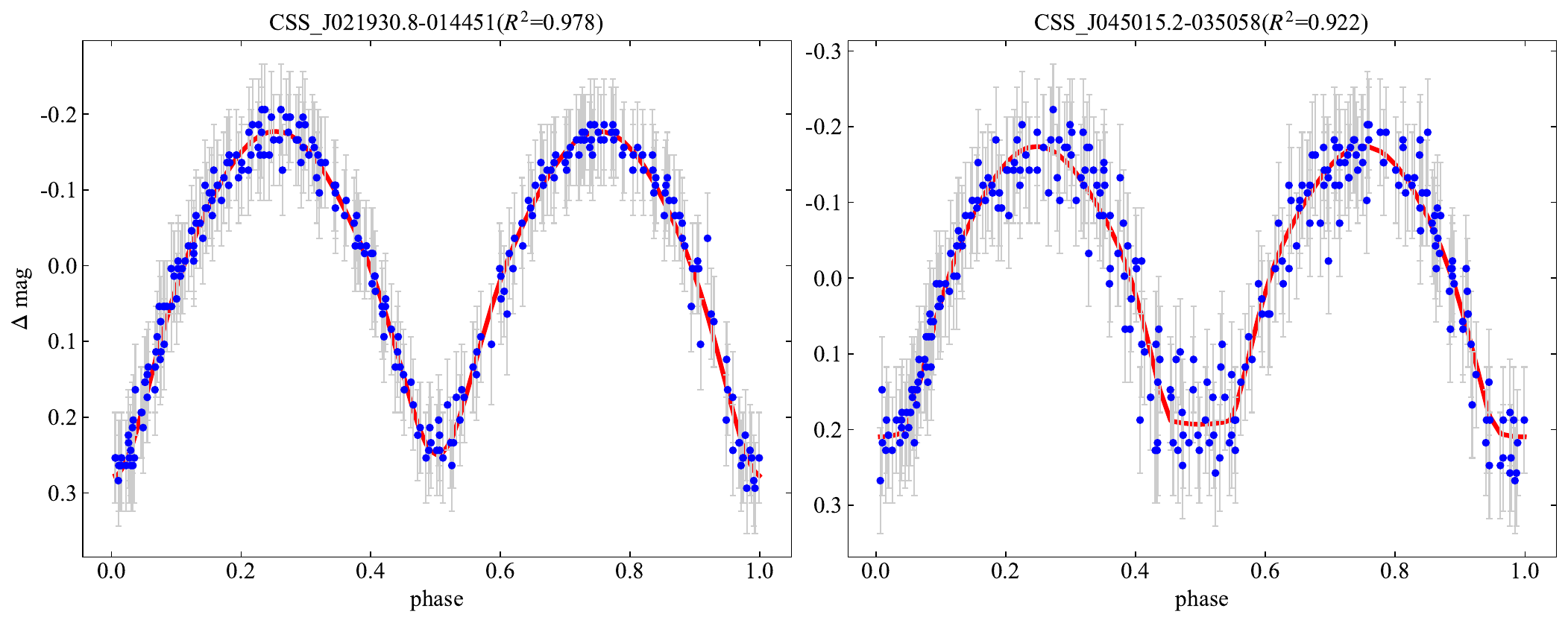}
    \caption{The fitting results for CSS\_J021930.8$-$014451 and CSS\_J045015.2$-$035058, where the blue dot and gray line show the observed light curve with photometric error, and the red line indicates the synthetic light curve generated by the input of parameters derived from the proposed method into the PHOEBE program.}
    \label{fig: r2_9}
\end{figure*}

\begin{figure*}
	\centering
	\includegraphics[scale=0.5]{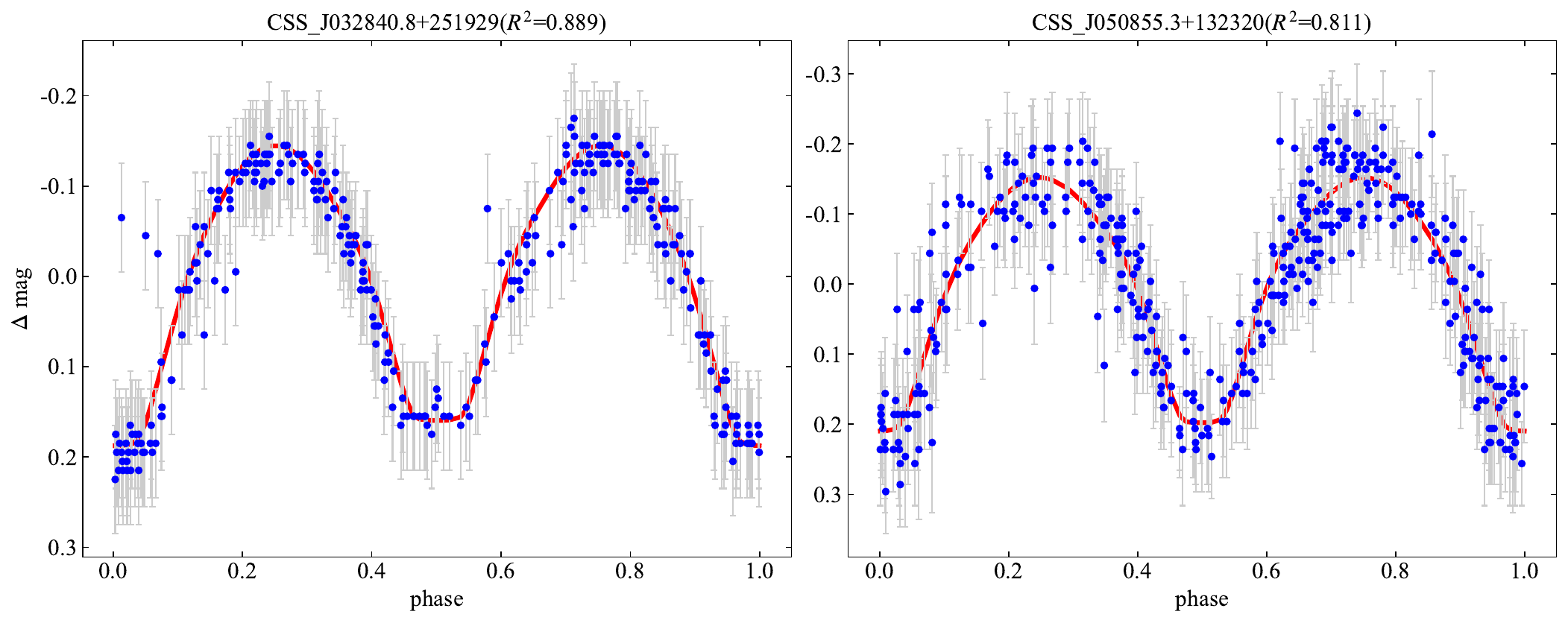}
    \caption{The fitting results for CSS\_J032840.8+251929 and CSS\_J050855.3+132320, where the blue dot and gray line show the observed light curve with photometric error, and the red line indicates the synthetic light curve generated by the input of parameters derived from the proposed method into the PHOEBE program.}
    \label{fig: r2_8}
\end{figure*}

\begin{figure*}
	\centering
	\includegraphics[scale=0.5]{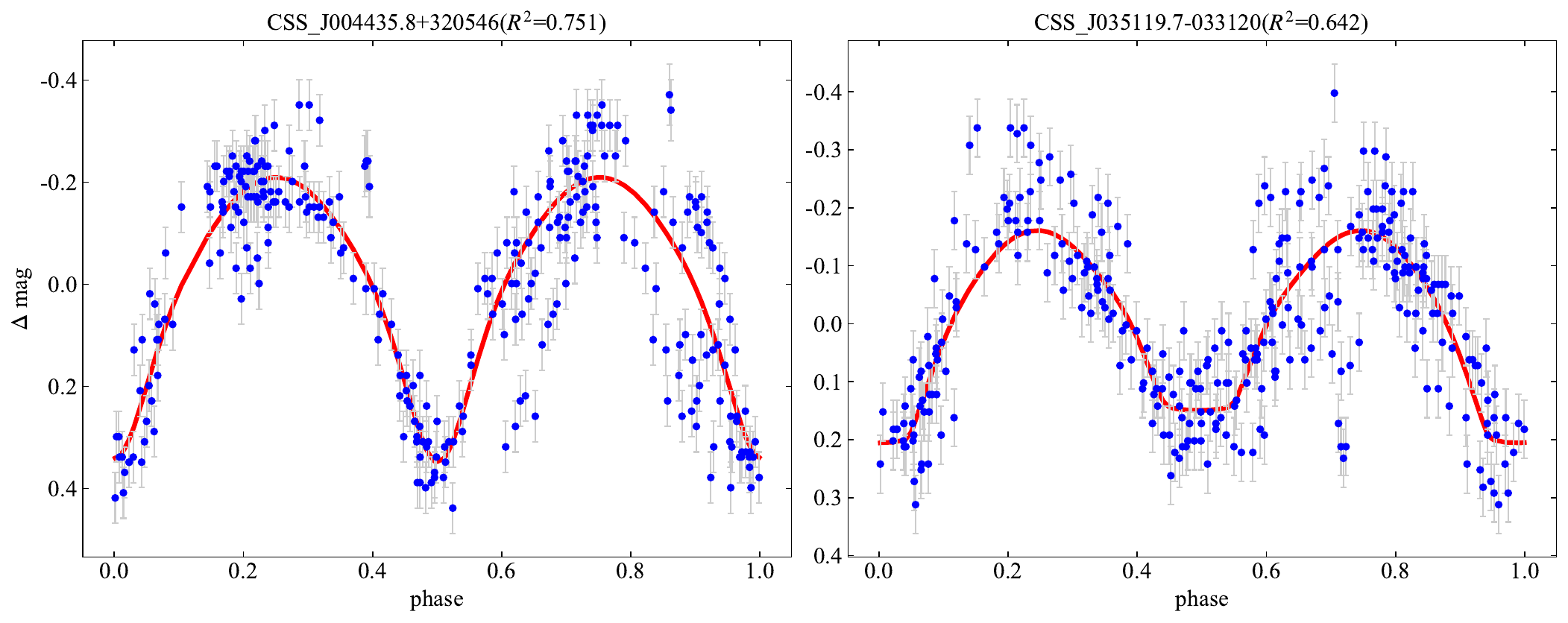}
    \caption{The fitting results for CSS\_J004435.8+320546 and CSS\_J035119.7$-$033120, where the blue dot and gray line show the observed light curve with photometric error, and the red line indicates the synthetic light curve generated by the input of parameters derived from the proposed method into the PHOEBE program.}
    \label{fig: r2_6}
\end{figure*}

\begin{figure*}
	\centering
	\includegraphics[scale=0.6]{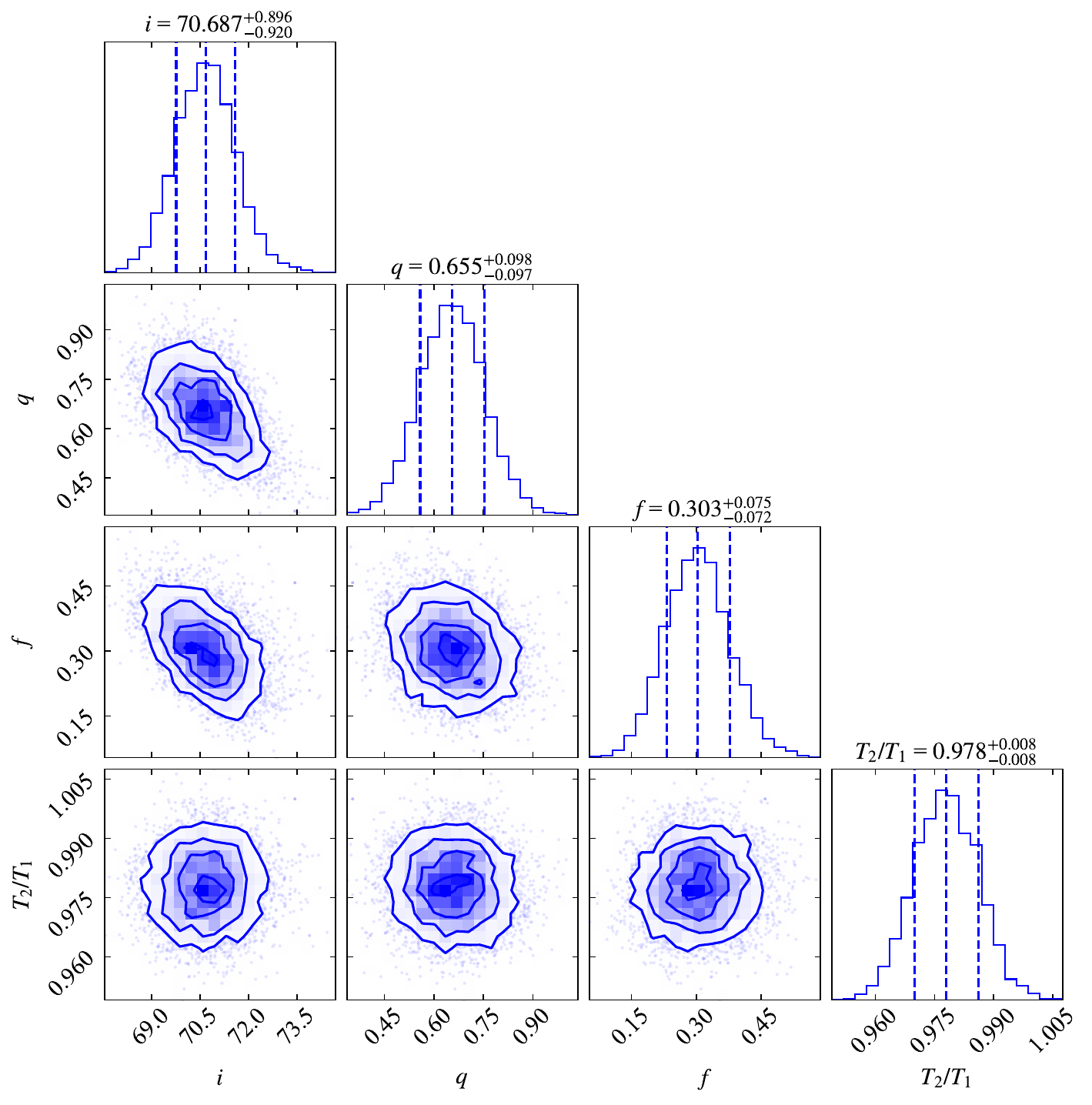}
    \caption{The posterior distribution of the 4 parameters ($i$, $q$, $f$ and $T_2/T_1$) of CSS\_J021930.8$-$014451 derived by the NNHMC method. The contours from the inside out enclose the 68, 95 and 99 percentiles of the total probability in the six off-diagonal panels. The three vertical dashed lines in the four diagonal panels represent the lower error at the 16 percentile, the median at the 50 percentile, and the upper error at the 84 percentile of the distribution, respectively.}
    \label{fig: CSS_corner}
\end{figure*}

The obtained 5172 CB parameters were input into the PHOEBE program to regenerate the light curve, and the $R^2$ and the root mean square error (RMSE) between the light curve and the observed light curve were calculated. The distribution of $R^2$ and RMSE is shown in Figure~\ref{fig: r2_rmse_distribution}. All $R^2$ values were greater than 0.6 and were mainly concentrated around 0.9. The RMSE is mainly concentrated around 0.05 mag, which is basically consistent with the photometric uncertainty of CBs in the CSS. Next, the parameter estimates for different $R^2$ values were shown. Two targets with $R^2$ values greater than 0.9 (CSS\_J021930.8$-$014451 and CSS\_J045015.2$-$035058) were randomly selected from 5172 CBs, and the results are shown in Figure~\ref{fig: r2_9}. Two targets with $R^2$ values between 0.8 and 0.9 were selected (CSS\_J032840.8+251929 and CSS\_J050855.3+132320), and the results are shown in Figure~\ref{fig: r2_8}. A target with an $R^2$ value between 0.7 and 0.8 (CSS\_J004435.8+320546) and a target with an $R^2$ value between 0.6 and 0.7 (CSS\_J035119.7$-$033120) were selected, and the results are shown in Figure~\ref{fig: r2_6}. In Figure~\ref{fig: r2_9}, Figure~\ref{fig: r2_8} and Figure~\ref{fig: r2_6}, the blue dot and gray line show the observed light curve with photometric error, and the red line indicates the synthetic light curve generated by the input of parameters derived from the proposed method into the PHOEBE program. It can be seen that the more concentrated and continuous the data points of the observed light curve, the better the fitting effect. As the data points of the observed light curve continue to disperse, the degree of fit decreases. Figure~\ref{fig: CSS_corner} shows the posterior distribution of parameters for one of the targets (CSS\_J021930.8$-$014451), from which it can be seen that the parameter distribution obtained by the proposed method basically conforms to the Gaussian distribution. Through the posterior distribution of parameters, the uncertainty of each parameter can be obtained. The parameters and uncertainty results of the six targets obtained by the proposed method are shown in Table~\ref{Tab:2}.

\begin{figure}
	\centering
	\includegraphics[scale=0.6]{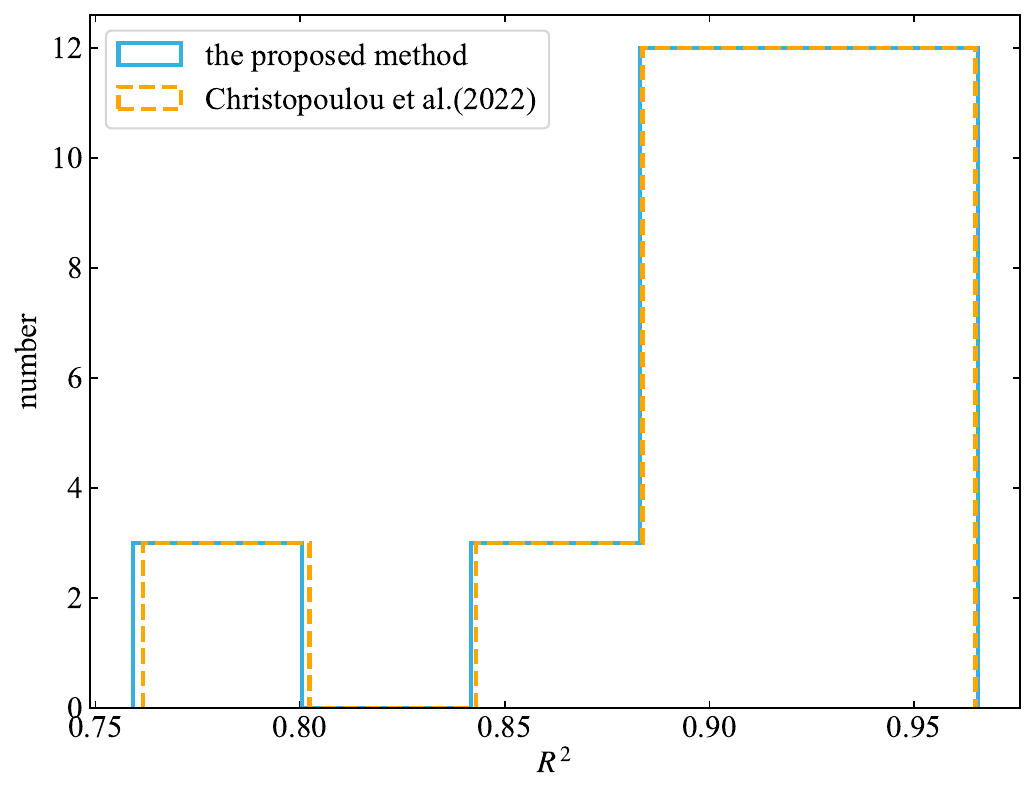}
    \caption{Comparison of the $R^2$ values between the parameters of 30 CBs in CSS derived by \cite{2022MNRAS.512.1244C} and the parameters re-derived using the method proposed in this paper, where the blue solid bar shows the $R^2$ distribution curve obtained by the proposed method, and the orange dashed bar indicates the $R^2$ distribution curve obtained by \cite{2022MNRAS.512.1244C}.}
    \label{fig: hmc_Christopoulou_r2}
\end{figure}

\cite{2022MNRAS.512.1244C} used the PHOEBE program combined with the grid search method to obtain the parameters of 30 eclipsed CBs from the CSS. Similarly, this study calculated their respective $R^2$, and the $R^2$ distribution curve is shown in Figure~\ref{fig: hmc_Christopoulou_r2}. The blue solid bar shows the $R^2$ distribution curve obtained by the proposed method, and the orange dashed bar indicates the $R^2$ distribution curve obtained by \cite{2022MNRAS.512.1244C}. Doing hypothesis testing \citep{raftery1995hypothesis} on the two distributions, the test results showed that the two distributions belonged to the same distribution, and the results were basically the same.

\begin{figure}
	\centering
	\includegraphics[scale=0.6]{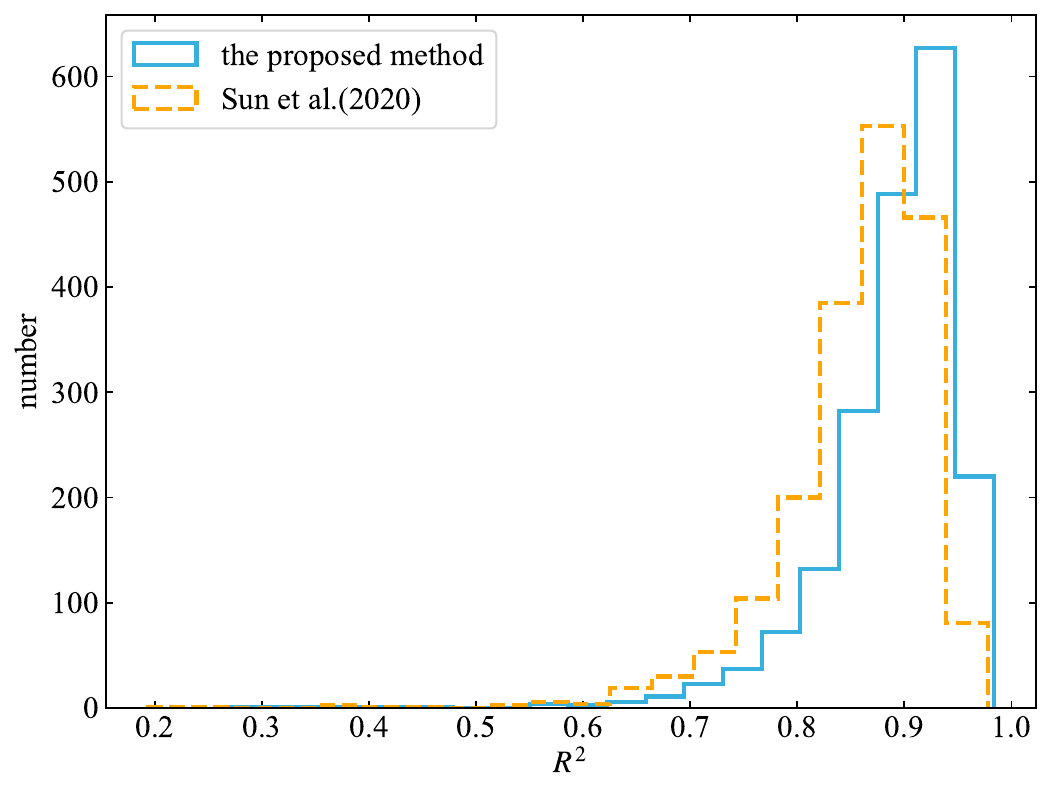}
    \caption{Comparison of the $R^2$ values between the parameters of 2335 CBs in CSS derived by \cite{2020ApJS..247...50S} and the parameters re-derived using the method proposed in this paper, where the blue solid bar shows the $R^2$ distribution curve obtained by the proposed method, and the orange dashed bar indicates the $R^2$ distribution curve obtained by \cite{2020ApJS..247...50S}.}
    \label{fig: hmc_Sun_r2}
\end{figure}

\cite{2020ApJS..247...50S} used the WD program and the q-search method to get the parameters of 2335 CBs from the CSS. In this paper, the proposed method was used to derive these CBs, and the $R^2$ between the obtained parameters by \cite{2020ApJS..247...50S}. and the obtained parameters by the proposed method and the observed light curve was calculated respectively, and the results are shown in Figure~\ref{fig: hmc_Sun_r2}. The blue solid bar shows the $R^2$ distribution curve obtained by the proposed method, and the orange dashed bar indicates the $R^2$ distribution curve obtained by \cite{2020ApJS..247...50S}. Doing hypothesis testing \citep{raftery1995hypothesis} on the two distributions, the test results showed that the two distributions do not belong to the same distribution, which was a difference in methodology. However, from the results, the parameters obtained by the proposed method have a better fit to the observed light curve.

The purpose of crossing with Gaia was to obtain an initial estimate of the primary star's temperature, which serves as a starting point for the HMC algorithm's exploration. The more accurate the initial estimate of the primary star's temperature, the less uncertainty there is in the final estimate. By removing systems with temperature differences greater than 400 K, a relatively accurate initial estimate of the primary star's temperature is ensured. Some binarity can be identified in the future using the composite spectrum or astrometry \citep{2018A&A...616A...8A} to obtain a more accurate estimate of the primary star's temperature and reduce the uncertainty in the final estimate of the primary star's temperature.

\begin{deluxetable*}{llllll}
\tablenum{1}
\tablecaption{Contents of Catalog. \label{Tab:1}}
\tablewidth{0pt}
\tablehead{
\colhead{Num}     &  \colhead{Column}                  &  \colhead{Units}        &  \colhead{Explanations}
}
\startdata
1       &  ID                      &               &  CSS ID\\
2       &  R.A.                      &  deg     &  R.A.(J2000)\\
3       &  decl.                     &  deg     &  Decl.(J2000)\\
4       &  Period                  &  day          &  Orbital period\\
5       &  $T_{\rm{init}}$              &  K            &  Initial effective temperature of primary star from Gaia DR2\\
6       &  $T_1$                   &  K            &  Effective temperature\\ 
        &                          &               &  of primary star\\
7       &  $\sigma_{T_1}$          &  K            &  Uncertainty in $T_1$\\ 
8       &  $i$                  &  deg     &  Inclination angle\\
9       &  $\sigma_{i}$         &  deg     &  Uncertainty in $i$\\
10      &  $q$                     &               &  Mass ratio\\
11      &  $\sigma_q$              &               &  Uncertainty in $q$\\
12      &  $T_2/T_1$               &               &  Temperature ratio\\
13      &  $\sigma_{T_2/T_1}$      &               &  Uncertainty in $T_2/T_1$\\
14      &  $f$                     &               &  Fill-out factor $f=(\Omega-\Omega_{in})/(\Omega_{out}-\Omega_{in})$,\\
        &                          &               &  where $\Omega_{in}$ and $\Omega_{out}$ are the modified potential\\
        &                          &               &  of the inner and the outer Lagrangian points,\\
        &                          &               &  respectively.\\
15      &  $\sigma_f$              &               &  Uncertainty in $f$\\
16      &  $\Omega$                &               &  Surface potential $\Omega=\Omega_1=\Omega_2$\\ 
17      &  $\sigma_{\Omega}$       &               &  Uncertainty in $\Omega$\\ 
18      &  $r_1$                   &               &  Relative radius of\\
        &                          &               &  primary star\\
19      &  $\sigma_{r_1}$          &               &  Uncertainty in $r_1$\\
20      &  $r_2$                   &               &  Relative radius of\\
        &                          &               &  secondary star\\
21      &  $\sigma_{r_2}$          &               &  Uncertainty in $r_2$\\
22      &  $L_2/L_1$               &               &  Light ratio\\
23      &  $\sigma_{L_2/L_1}$      &               &  Uncertainty in $L_2/L_1$\\
24      &  $R^2$                   &               &  goodness-of-fit\\
\enddata
\tablecomments{This table is available in its entirety in machine-readable form.}
\end{deluxetable*}

\begin{deluxetable*}{lcccccl}
\tablenum{2}
\tablecaption{The parameters of the six new CSS CBs derived by NNHMC method. \label{Tab:2}}
\tablewidth{0pt}
\tablehead{
\colhead{ID} & \colhead{$i$} & \colhead{$q$} & \colhead{$f$} & \colhead{$T_2/T_1$}
}
\startdata
CSS\_J021930.8$-$014451  & $70.69\pm0.94$   	& $0.66\pm0.10$     & $0.30\pm0.07$  & $0.978\pm0.008$\\
CSS\_J045015.2$-$035058  & $86.27\pm0.96$   	& $0.18\pm0.01$ 	& $0.69\pm0.09$  & $0.995\pm0.008$\\
CSS\_J032840.8+251929    & $78.61\pm2.68$       & $0.14\pm0.01$ 	& $0.60\pm0.10$  & $0.979\pm0.009$\\
CSS\_J050855.3+132320    & $79.22\pm2.79$   	& $0.19\pm0.02$ 	& $0.30\pm0.09$  & $0.991\pm0.008$\\
CSS\_J004435.8+320546    & $75.36\pm0.61$  	    & $0.78\pm0.10$ 	& $0.33\pm0.06$  & $1.004\pm0.006$\\
CSS\_J035119.7$-$033120  & $87.41\pm2.09$   	& $0.17\pm0.01$ 	& $0.71\pm0.09$  & $0.952\pm0.007$\\ 
\enddata
\end{deluxetable*}

\section{Conclusion}\label{sec:Conclusion}

In this paper, a new rapid derivation method for light curves based on the NN model combined with the HMC algorithm was developed and applied to the CBs from the CSS. The NNHMC method was used to quickly derive the parameters of the CBs. Firstly, an NN model of parameter to light curve was established to generate synthetic light curves efficiently. Then, the HMC algorithm was employed to rapidly derive the observed light curves. Finally, the posterior distribution of parameters was obtained, and its median and error were calculated. On a computer with 16GB random-access memory and a 3.40 GHz 16-core CPU, the average time required to generate a synthetic light curve with 100 phase points decreased from 5s to $0.001$s. The precision of generating synthetic light curves using the NN model was around three parts per ten thousand, which significantly exceeds ground-based survey measurement accuracy while meeting their required level of precision simultaneously. The NNHMC method possesses the capability to simultaneously derive parameters for a vast number of targets, thereby rendering it an exceedingly efficient tool for quickly deriving parameters in forthcoming sky surveys encompassing extensive samples of CBs.

Additionally, the NNHMC method was validated by randomly selecting 5000 synthetic light curves in the test set and adding Gaussian noise with a $\sigma$ of $0.01$ magnitude. The results show that the $R^2$ distribution obtained by the NNHMC method was basically the same as those obtained by \cite{2022AJ....164..200D}, but the time to compute a light curve was shortened by 2 orders of magnitude on the same computational platform. Notably, the more light curves, the higher the efficiency.

Simultaneously, the NNHMC method was used to derive the parameters of 19,104 CBs in the CSS, among which 5172 with an inclination greater than $70^\circ$ and a temperature difference less than 400 K were identified. These derived parameters were then input into the PHOEBE program, and the $R^2$ value between the generated light curve and the observed light curve was utilized as an evaluation index for assessing the effectiveness of these derived parameters. The $R^2$ distribution of both methods was found to be largely consistent when compared to the 30 CBs presented in \cite{2022MNRAS.512.1244C}. Additionally, data fitting results are shown for different ranges of $R^2$ values: above 0.9, between 0.8 and 0.9, between 0.7 and 0.8, and between 0.6 and 0.7. The $R^2$ of 5172 CBs was greater than 0.6, of which 4,219 have a $R^2$ greater than 0.8 and 1,821 have a $R^2$ greater than 0.9. The results show that the more concentrated and continuous the data points of the observed light curve, the better the fitting effect.

\begin{acknowledgments}

This work is supported by the Natural Science Foundation of China (Nos.\ 12103088, 12303106). We are very grateful for the data released by the CSS. The funding for the CSS is provided by the National Aeronautics and Space Administration through Grant No.\ NNG05GF22G, administered by the Science Mission Directorate Near-Earth Objects Observations Program. The Catalina Real-Time Transient Survey receives support from the US National Science Foundation through grants AST-0909182, AST-1313422, AST-1413600, and AST-1518308.

\end{acknowledgments}

\bibliography{sample631}{}
\bibliographystyle{aasjournal}

\end{document}